# Optimization of InAs/InAlGaAs/InP quantum dots for O-band laser applications

Anna Penkała [a,*], Paweł Podemski [a], Vinayakrishna Joshi [b], Vitalii Sichkovskyi [b], Johann P. Reithmaier [b] and Grzegorz Sęk [a]

[a] *Department of Experimental Physics, Faculty of Fundamental Problems of Technology, Wrocław University of Science and Technology, Wybrzeże Stanisława Wyspiańskiego 27, 50-370 Wrocław, Poland*

[b] *Technological Physics, Institute of Nanostructure Technologies and Analytics, CINSaT, University of Kassel, 34132 Kassel, Germany*

## Abstract

The O-band telecommunication spectral range combines minimal chromatic dispersion with low transmission loss, making it highly suitable for optical communications. Semiconductor telecom lasers utilizing low-dimensional active regions, such as quantum dots, offer low threshold power and high thermal stability. However, long-wavelength quantum dot emitters often suffer from thermal carrier escape through excited states when the energy separation between the ground and excited states is insufficient. Here we demonstrate the optimization of InAs quantum dot emitters grown on InP by molecular-beam epitaxy for O-band laser active regions. By optimizing the design for better carrier capture efficiency and tuning the emission wavelength to the O-band, we show ways of improving the carrier confinement. Absorption spectroscopy on the finalized structure confirms substantial energy separation between the quantum dot ground and excited states, effectively suppressing thermal carrier escape even at elevated temperatures. These results highlight critical design for improving carrier capture and retention in long-wavelength quantum dot lasers.

## Introduction

The optical fibers, formed from flexible glass and extensively used in telecommunication, have spectral properties that resulted in defining of a few telecommunication bands, where certain wavelengths are favored for the signal transmission. One of them, O-band, spanning spectrally from 1260 to 1360 nm, provides low transmission losses, offering at the same time very weak chromatic dispersion, limiting undesirable effects of pulse broadening and chirping. Semiconductor lasers operating at these wavelengths can take advantage of low-dimensional active region composed of quantum wells or quantum dots (QDs). Especially the latter can provide low threshold power [1,2], high temperature stability [3-5] due to the QDs' inherent discrete energy level structure and broad gain due to intrinsic active region inhomogeneity and the resulting device tunability [6]. These advantages of QDs can, however, become less prominent when the energy levels' spacing approaches carriers' thermal energy. When the energy levels are insufficiently separated the probability of carrier loss increase instead of contributing to the lasing. This issue becomes especially important in QDs emitting at longer wavelengths (e.g. at O-band and C-band) as these QDs are typically larger, resulting in a lower separation of energy levels.

QDs designed for O-band emission and grown on an InP substrate are more challenging when compared to well-established GaAs-based systems [7-9]. Still, the possibility of significantly higher modal gain with InP substrate makes InAs/InP QD system very appealing [10]. The InP-based QD system also offers ultra-fast charge carrier dynamics with temperature-stable high modulation bandwidths [11,12]. Self-assembled QDs formed from InAs on an InP substrate is driven by lower lattice mismatch (approx. 3%) favoring naturally longer wavelengths – event through the tele-com C-band spectral range. However, the growth engineering, like strain control or use of intermediate materials (e.g. InAlGaAs) allows also for spectral matching of InP-based QDs to telecom O-band [13,10].

[*] Corresponding author. E-mail address: anna.penkala@pwr.edu.pl

Here, we look at the optical properties of series of structures with InAs quantum dots grown (indirectly) on an InP substrate by molecular-beam epitaxy (MBE). The quantum dots are designed as an active region of lasers operating in the telecom O-band. The active region should have limited carrier losses so we grew a few structures differing in the composition of a barrier surrounding quantum dots and we observed the optimization steps influence on the QD emission wavelength. At the same time we looked for QD excited states spectral features, as one of the easiest escape routes (i.e. requiring low thermal energy) are higher states confined within QDs.

## Materials and methods

The investigated structures were grown on sapphire-doped InP substrate by MBE in a three-chamber setup in a general layout presented in Fig. 1. Directly on the substrate there was deposited 100-nm-thick InP buffer layer, followed by the InAlGaAs barrier of the same thickness below and above the active region.

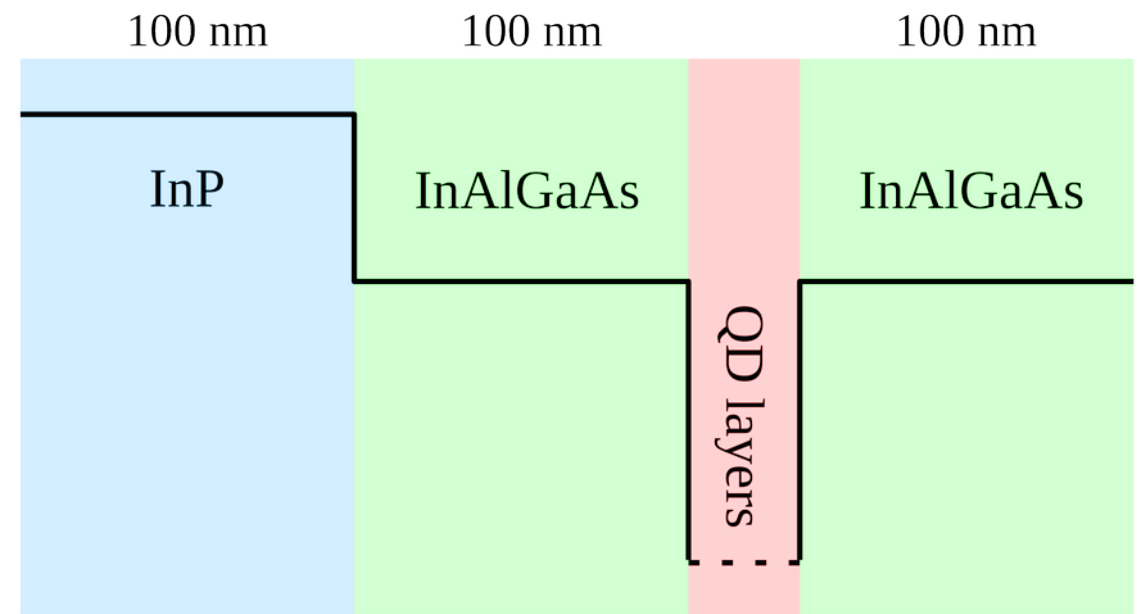

Fig. 1. General layout of QD structures with schematic conduction band edge indicating the carrier confinement

The growth of QDs was preceded by submonolayer of GaAs (thickness between 0,4 and 0,5 nm) to make the uniform growth of QDs more favorable. Next the QD InAs material was deposited, forming dots in the Stranski-Krastanov growth mode. The QD layer was covered with InAlGaAs barrier with the same composition as the bottom one (see Table 1). The number of QD layers, their thickness and the barrier composition varied between the structures as presented in Table 1. Consecutive QD layers in sample C were separated by 20 nm of InAlGaAs.

Table 1
Differences between the investigated QD structures. ML stands for monolayers

| Sample | Barrier | QD layer thickness | Number of QD layers |
|---|---|---|---|
| A | $In_{0.528}Al_{0.238}Ga_{0.234}As$ | 3.0 ML | 1 |
| B | $In_{0.530}Al_{0.350}Ga_{0.120}As$ | 3.4 ML | 1 |
| C | $In_{0.530}Al_{0.350}Ga_{0.120}As$ | 4.5 ML | 10 |

Emission from the structures was investigated in photoluminescence (PL) setup. The samples were cooled-down to 5 K using continuous-flow microscopy cryostat to minimize the impact of thermal effects on the optical response. As an excitation source we used 639 nm line of a CW semiconductor laser and the emission was recorded with a nitrogen-cooled InGaAs linear array detector coupled to 1000-mm-focal-length monochromator. For the photoreflectance (PR) measurements we used a lock-in detection method [14] with the modulation beam provided from a frequency-doubled neodymium-doped yttrium aluminum garnet (Nd:YAG) laser (532 nm). The PR signal was registered by a thermoelectrically cooled long-wavelength InGaAs photodiode connected to a 300-mm-focal-length monochromator. The photoluminescence excitation spectra (PLE) were obtained using similar cooling system as in PL measurements. As the tunable excitation we used a titanium-sapphire laser (pulsed operation at 76 MHz) feeding an optical parametric oscillator with the detection similar to the PL measurement but with the shorter focal length of a monochromator (500 mm).

**Results and discussion**

The first investigated structure (sample A) was grown with $In_{0.528}Al_{0.238}Ga_{0.234}As$ barriers and 3 monolayers (ML) of InAs, being a dot-forming material. Its emission spectra for various excitation power values in the range of 100 nW – 2 mW is shown in Fig. 2.

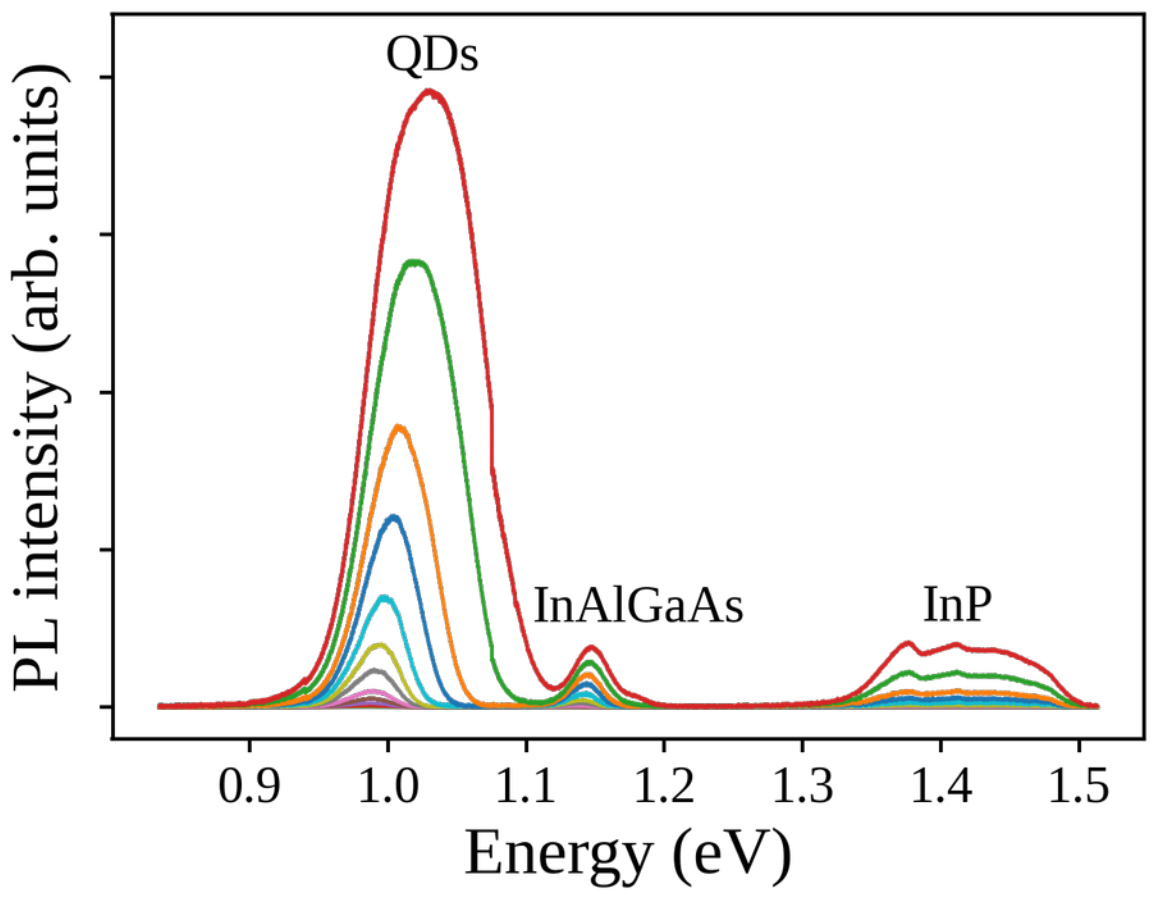

Fig. 2. Low-temperature PL spectra from structure A for different excitation power values (100 nW – 2 mW)

The emission from QDs is centered at 0.98 eV (1260 nm) for low excitation power and shifts to 1.03 eV (1205 nm) with excitation increase. Even at cryogenic temperatures it overlaps with the intended O-band spectral range; for higher temperatures the spectral matching will be further improved due to the temperature-driven bandgap shrinkage. This kind of slight energy shift with the excitation power is not typically expected in self-assembled QDs. At higher temperatures, where the thermal energy allows for carrier redistribution within the QD ensemble, it is more likely. Then, at lower excitation the carriers tend to occupy QDs with lowes ground state energy. Increase of the excitation power provides more carriers which leads to filling also the states of the higher-energy QDs. In this case, however, at temperature of 5 K, the carriers' redistribution effect is expected to much less significant (or completely negligible, depending on the dots density and their spatial separation as well as availability of the carrier transfer channels between the dots), so QD emission should not shift with the excitation power increase. Another expected spectral feature would be the state-filling effect where increase in the excitation causes a sequential occupation of the higher energy states; if their separation is smaller than the inhomogeneous broadening then the overall influence on the emission band is blue shift. But this is usually observed as saturation of the ground-state emission with following increase in the intensity of higher-energy PL maxima. But this is not observed in our case. However, if this is observed it would indicate on the very small separation between successive QD energy levels, which is unfavorable from the device characteristics point of view. In Fig. 2. we also observe emission at higher energy (approx. 1.15 eV), which comes from the InAlGaAs barrier. The spectrally-wide emission at 1.42 eV is related to the InP material and the broadening is related to the growth imperfections and the differences between the sapphire-doped InP substrate and MBE-grown InP part of the structure, where the defect states in InP result in the low-energy emission maxima (e.g. bound excitons, donor-acceptor transitions) [15,16].

For the next structure (sample B), we increased the InAlGaAs barrier energy by incorporating greater fraction of Al (up to 0.350). At the same time we slightly increased the QD material thickness (from 3.0 to 3.4 ML). Increased potential barrier should decrease the probability for carriers' escape from QDs, which is crucial with respect of the lasing threshold and its sensitivity to temperature (characteristic $T_0$ temperature). Fig. 3 shows the emission spectra from sample B for the excitation power values of 500 nW – 2 mW.

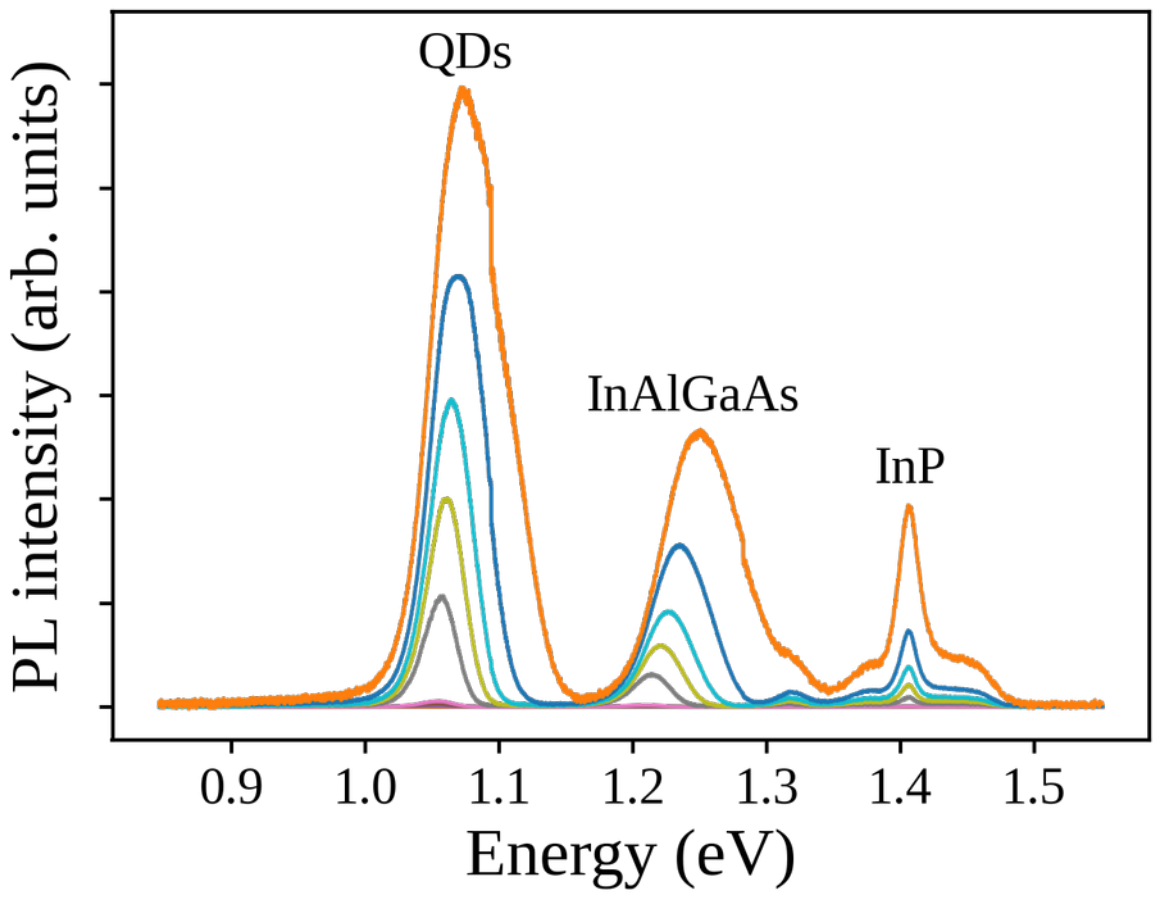

Fig. 3. Low-temperature PL spectra from structure B
for different excitation power values (500 nW – 2 mW)

The change of the InAlGaAs composition increased its energy, as expected, from 1.15 eV for structure A to above 1.20 eV for structure B. We observe also a noticeable energy shift of the barrier emission energy (from 1.21 to 1.25 eV) with excitation power, which may suggest some composition non-uniformity or defect states within the barrier material. At the same time the QD emission shifted from about 1.05 eV (1180 nm) to 1.08 eV (1150 nm), i.e. is slightly higher than observed for sample A. It is most likely a result of the higher Al content in the InAlGaAs barrier which, during MBE growth, can be incorporated to the QDs (grown directly on InAlGaAs) increasing their ground state energy.

To keep the higher confinement barriers and at the same time to shift back the QD emission towards the O-band, for structure C we increased the nominal thickness of the QD material during the MBE growth (from 3.4 to 4.5 ML) and kept the InAlGaAs layers compositions as for sample B. We also decided to include at this step a larger number of QD layers which makes the C structure closer to an active region design of an operational laser device. The respective emission spectra versus excitation power are presented in Fig. 4.

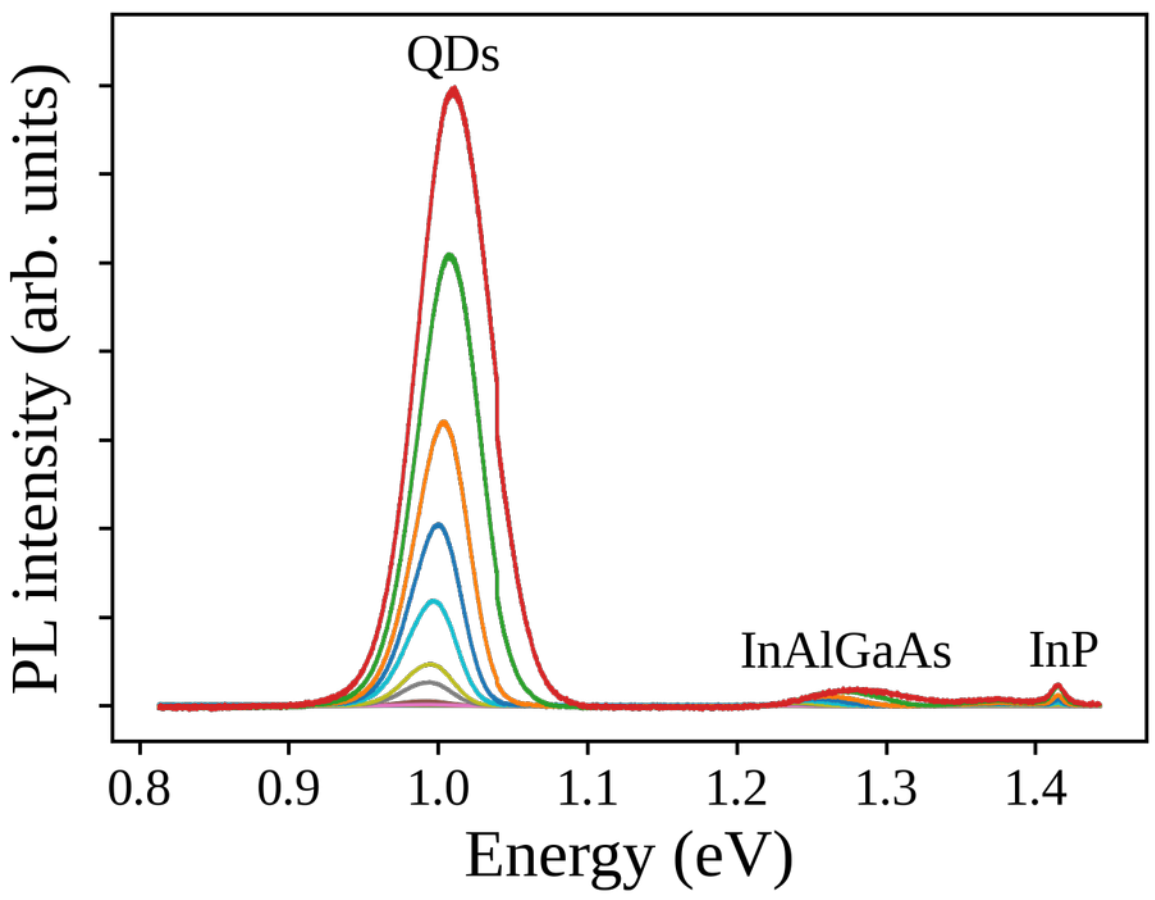

Fig. 4. Low-temperature PL spectra from structure C
for different excitation power values (200 nW - 10 mW)

InAlGaAs barrier PL peak energy in the C structure remained indeed similar as for sample B, 1.23 to 1.28 eV, with the excitation increase. At the same time, QD emission is redshifted as intended: changes from 0.99 eV (1250 nm) to 1.01 eV (1230 nm), depending on the excitation power. At the same time, the increased number of QD layers is also reflected in stronger QD emission, relative to InAlGaAs barrier and InP substrate PL peaks, as expected. It is worth nothing that although the QD PL bands of samples A and C are similar, the energy shift with the excitation power is significantly different: 20 meV for structure C versus 50 meV for A.

We provide more quantitative information in Fig. 5, which shows QD energy and full width at half maximum (FWHM) dependencies on the excitation power.

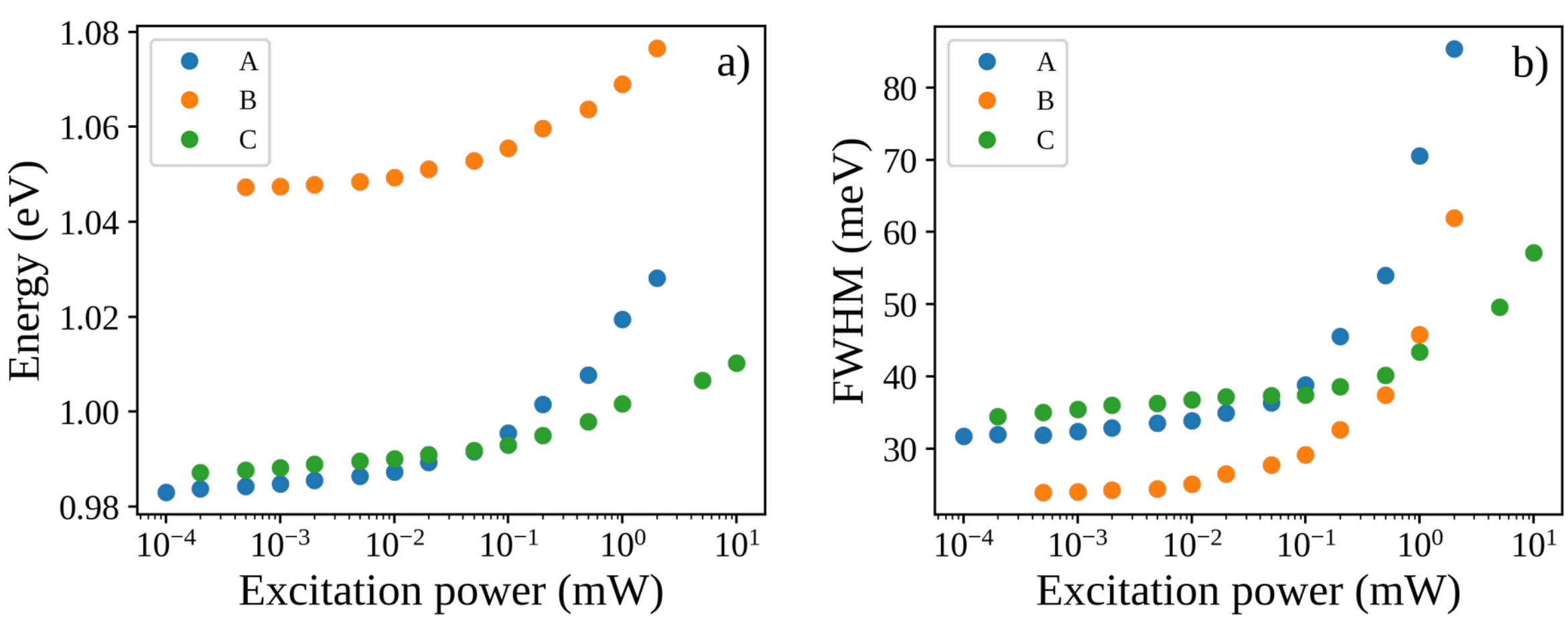

Fig. 5. Low-temperature QD emission energy a) and FWHM b) dependencies on the excitation power

In Fig. 5 a) we see that the increase of the QD material thickness during the MBE growth (from 3.4 for structure B to 4.5 ML for structure C) indeed moved the QD emission energy back below 1 eV. The emission reached the level of the QD emission in structure A, while keeping the increased In-AlGaAs barrier energy. We clearly observe lower QD emission energy shift with the excitation power for structure C, as was discussed above. As this energy shift is most likely related to the low energy separation between consecutive QD energy levels, the increase of the excitation power should also result in broadening of the observed QD emission (covering ground and excited QD states). In Fig. 5 b) we show that the QD emission FWHM follows the observed energy shift for all the structures, increasing approximately 2-3 times for structures A and B for higher excitation power values. However, both, the energy shift and the FWHM dependencies for structure C, are significantly weaker, what may indicate on lower QD energy level separation. This issue is further elaborated below yet.

To provide more insight into the energy structure we used two absorption-like techniques – PR and PLE. In Fig. 6, we present the room-temperature PR, PL, and PLE spectra for structure A.

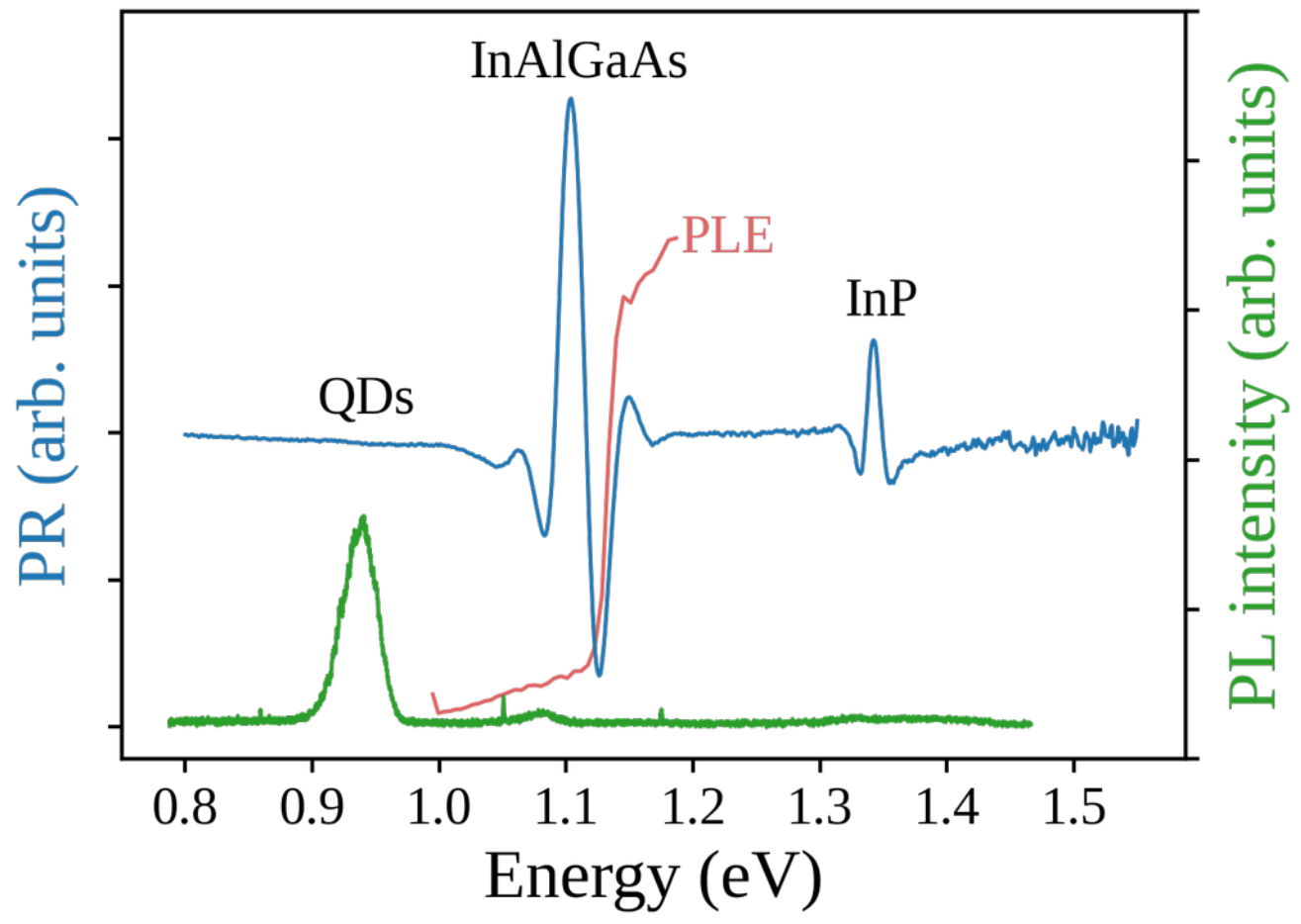

Fig. 6. Room-temperature PR spectrum for structure A, together with PL and PLE spectra (measured at 5 K and shifted to the room temperature)

To prevent the temperature-induced carriers' Fermi distribution change and the temperature impact on the distribution of carriers among the QDs, we measured PL and PLE spectra at low temperature (5 K) and then shifted them to the room temperature (by 50 meV) to compare them directly with room-temperature PR spectra. The strongest response in PR spectrum comes from the InAlGaAs barrier, followed by the InP-related feature. The QD signal is very weak in PR, which is common and related to low absorption of QDs, as compared with absorption in the bulk-like layers. On the contrary, in PL the QD emission is clearly seen with much weaker emission intensity from the InAlGaAs barrier. When we look at the PLE spectrum (with the detection at the QD PL peak) we see almost no absorption at the energies below InAlGaAs barrier bandgap, as expected. There is only a slight slope, most likely arising from the continuum-of-states absorption [17,18] and presence of wetting layer states. The sharp edge marks the onset of the InAlGaAs barrier absorption – carriers are generated within InAlGaAs and then relax to QD states where they recombine radiatively.

The results for structure B, with increased InAlGaAs barrier energy, we show in Fig. 7. Modification of the InAlGaAs barrier composition has an impact on the PR response. The barrier-related feature is shifted to higher energies, as previously observed in emission spectra, but at the same time it becomes much broader and less prominent. As we can observe InP and even QD signal in PR spectrum, there is another strong feature at 1.26 eV. The change of the InAlGaAs barrier composition most likely introduced non-uniformity of the barrier material resulting in an additional PR response observed between InAlGaAs and InP energies. The PLE spectrum shows similar behavior above the QD emission energy with no distinct features and monotonous signal increase. There is, however, no sharp absorption edge observed, when approaching InAlGaAs energy (as identified in PR and PL spectra). It may confirm the assumption of the InAlGaAs non-uniformity and even suggest higher InAlGaAs bandgap energy (e.g. the energy of the strong feature at 1.26 eV), while the lower-energy InAlGaAs response in PR and PL spectra (at approximately 1.14 eV) could arise from defect-like states.

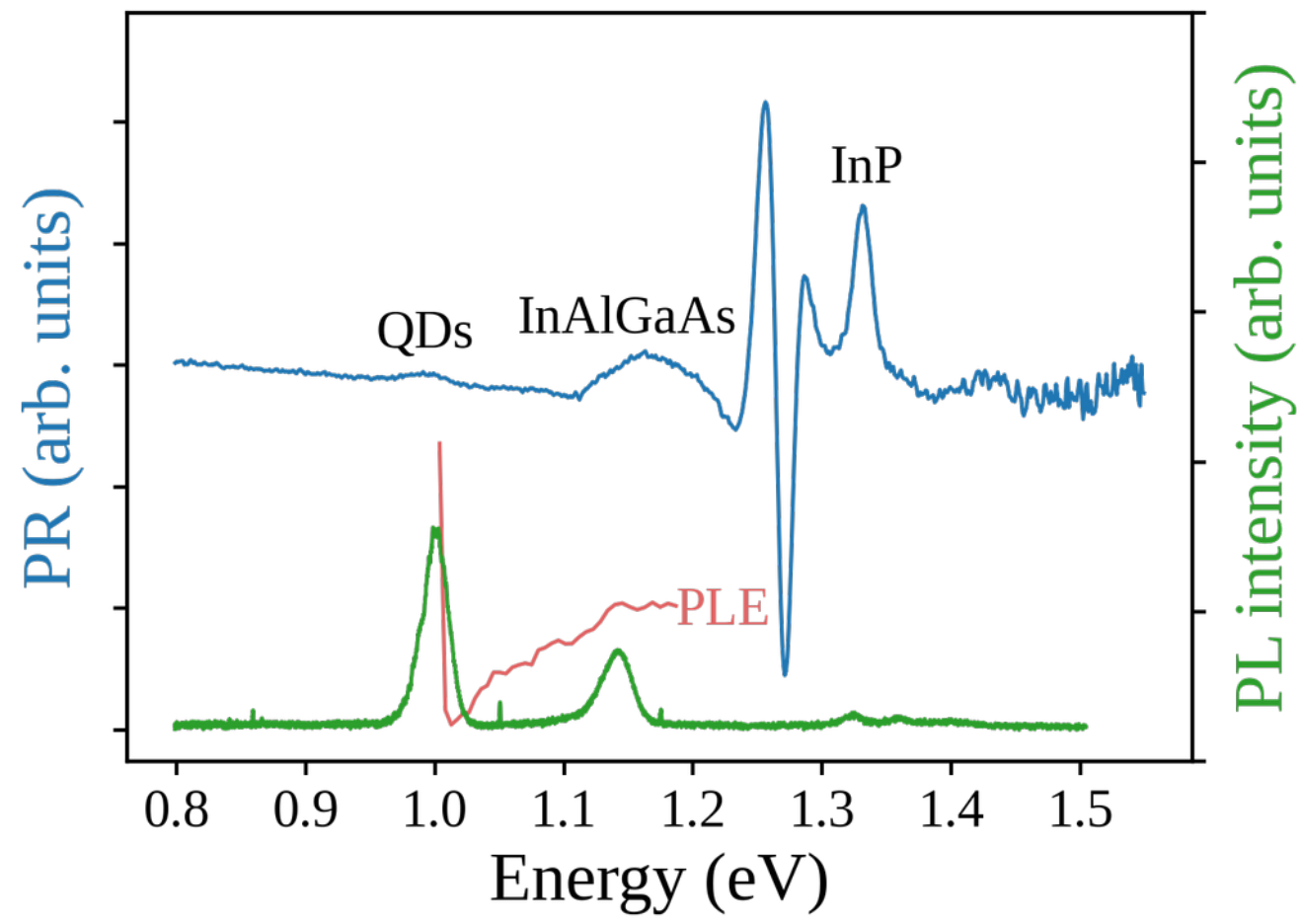

Fig. 7. Room-temperature PR spectrum for structure B, together with PL and PLE spectra (measured at 5 K and shifted to the room temperature)

The available tuning range in the PLE experiment, however, could not reach 1.26 eV allowing for the potential PLE absorption edge observation. There is also possibility that this strong feature in PR spectrum is related to the MBE-grown InP, impacted by the growth of the Al-rich InAlGaAs barrier. Nevertheless, the higher bandgap energy of InAlGaAs in structure B still provides improved barrier preventing escape of the carriers from QDs.

Finally, we look at the structure C, where high bandgap energy of InAlGaAs barrier is kept while the QD emission is tuned back to the lower-energy telecom range.

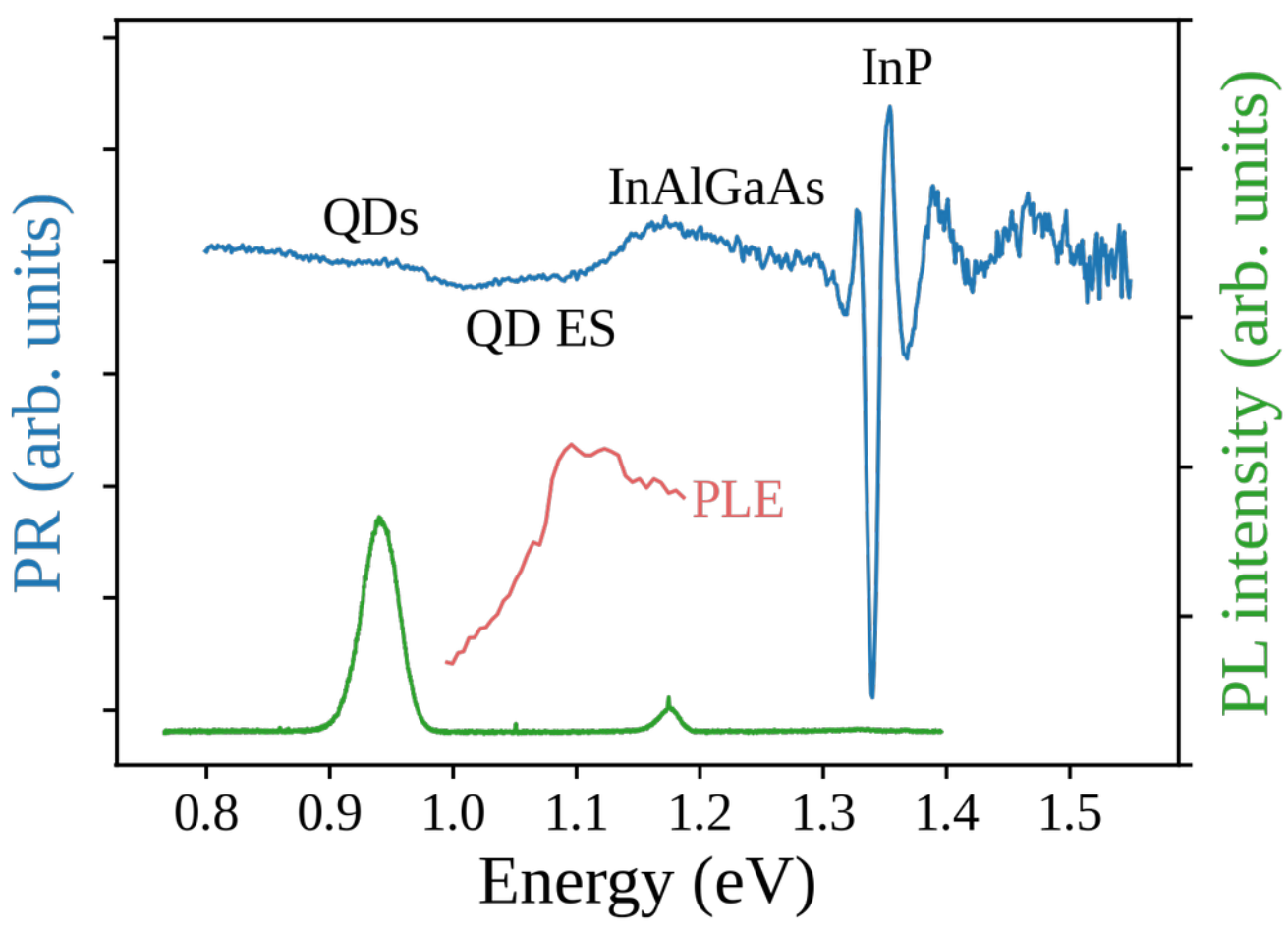

Fig. 8. Room-temperature PR spectrum for structure C, together with PL and PLE spectra (measured at 5 K and shifted to the room temperature)

In Fig. 8 we see again the broad PR response from the InAlGaAs barrier and weak signal from QDs – both in agreement with the PL spectrum. The high energy PR part is a little bit different, though. This time, the strong resonance is located at the energy corresponding to the InP substrate. It may indicate that in the structure B it was related to the disturbed purity of the MBE-grown InP. When we look at the PR spectrum in the energy range between QDs and InAlGaAs barrier, we notice additional features, suggesting existence of absorbing states with a carrier transfer to QD ground state. As in this spectral range we expect only higher QD states we marked it temporarily in Fig. 8 as QD excited states (QD ES). However, when we look at the PLE spectrum it no longer shows slow intensity increase only, but there appear maxima confirming absorption in this spectral range. When compared with observed PR features they support the identification of the observed features as higher QD states. We look more closely at the PR spectrum in this spectral range in Fig. 9 to resolve the particular transitions.

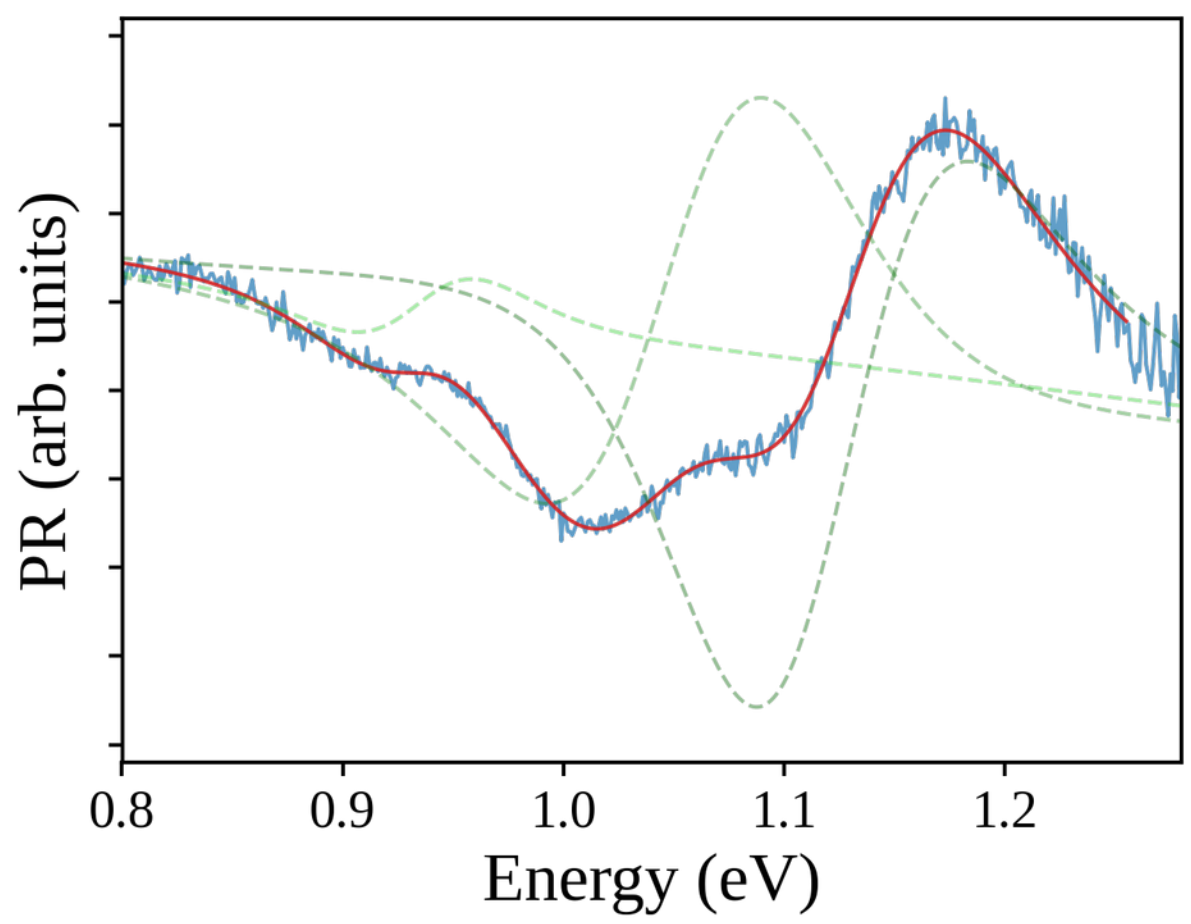

Fig. 9. Room-temperature PR spectrum for structure C in the spectral range of QD states.
The solid line represents PR signal fit, composed of three optical transitions (dashed lines).

For the structure C, the PR spectrum in the spectral range of QD states suggest observation of at least one QD excited state. For quantitative analysis we use the following formula to simulate the PR spectrum:

$$A+B\cdot E+\sum_{i=1}^{3}\frac{D_i}{\left(\left(E-E_i\right)^2+\Gamma_i^2\right)^{n/2}}\cdot\cos\left(\Theta_i-n\cdot\left(\frac{\pi}{2}-\arctan\left(\frac{E-E_i}{\Gamma_i}\right)\right)\right) \quad (1)$$

where $A$ and $B$ are parameters of a possible background, $E$ stands for the energy and $n$ represents the optical transition type. For the transitions in QDs at room temperature it takes on a value of 3 [14]. Each of the optical transitions is described by four parameters: intensity ($D_i$), energy ($E_i$), linewidth ($\Gamma_i$) and phase ($\Theta_i$). We tested the fit, starting from one transition and the lowest number of transitions, resulting in a satisfactory fitting, was three. We show the final PR fit (solid line) and the three optical transitions (dashed lines), from which it is composed (see Fig. 9). From this analysis we obtain three optical transition energy values: 0.94 eV, 1.06 eV and 1.12 eV. The lowest one we associate with the QD ground state transition, which also agrees with the QD emission shown in Fig. 8. The two higher transitions we attribute to QD excited states – they are both below InAlGaAs barrier energy and they agree well with the PLE maxima shown in Fig. 8. When we look at the energy difference between the QD ground and first excited state (120 meV) and between first and second QD excited states (60 meV) we see that these values correspond to thermal energies of 1400 K and 700 K, respectively. This provides very good separation between the consecutive QD energy levels, efficiently preventing carrier escape from QD ground state at room, or even elevated, temperatures, which is one of the critical parameters in laser structures.

## Conclusion

We presented optimization steps of a laser active region composed of InAs QDs grown by MBE on InP, with the intended emission in the O-band telecom spectral range. We followed the changes in the design of the structures – increase of the InAlGaAs barrier bandgap, for better carrier collection, and fixing the QD emission wavelength, while keeping the high barrier energy. For the final struc-ture we observed absorption in QD excited states with the QD high energy levels separation which should allow to minimize carrier losses from the QD ground state at temperatures significantly higher than the room temperature. The proposed design provides an InP-based platform for O-band QD lasers construction.

### Author contributions

**Anna Penkała:** Data Curation, Formal Analysis, Investigation, Writing – review & editing. **Paweł Podemski:** Conceptualization, Data Curation, Formal Analysis, Investigation, Supervision, Visualization, Writing – original draft. **Vinayakrishna Joshi:** Methodology, Resources, Writing – review & editing. **Vitalii Sichkovskyi:** Methodology, Resources, Writing – review & editing. **Johann P. Reithmaier:** Funding Acquisition, Project Administration, Methodology, Resources, Writing – review & editing. **Grzegorz Sęk:** Conceptualization, Formal Analysis, Funding Acquisition, Project Administration, Supervision, Writing – review & editing.

### Funding sources

This research did not receive any specific grant from funding agencies in the public, commercial, or not-for-profit sectors.

### Declaration of competing interest

The authors declare no conflicts of interest.

### Data availability

Data underlying the results presented in this paper are not publicly available at this time but may be obtained from the authors upon reasonable request.